\documentclass[11pt,english]{article}
\usepackage{graphicx}
\usepackage{amstext}
\usepackage{caption}
\usepackage{etoolbox}
\usepackage{makeidx}
\include{epsf}
\usepackage{amsfonts}
\usepackage{authblk} 
\usepackage{sectsty}
\usepackage{amsmath,amssymb,epsfig}
\usepackage{amscd}
\usepackage{amsthm}
\usepackage{mathrsfs}
\usepackage{dsfont}
\usepackage[applemac]{inputenc}
\usepackage[english]{babel}
\usepackage{enumitem} 
\usepackage[]{latexsym}
\usepackage{caption}
\usepackage{subfigure}
\usepackage{blindtext}
\usepackage{verbatim}
\usepackage{cancel}
\usepackage{color}
\usepackage{mathtools}

\newcommand*{\affaddr}[1]{#1}
\newcommand*{\affmark}[1][*]{\textsuperscript{#1}}

\newtheorem*{proof*}{Proof}

\DeclarePairedDelimiter\floor{\lfloor}{\rfloor}

\newcommand{\be}{\begin{equation}}

\newcommand{\ee}{\end{equation}}
\def\beqa{\begin{eqnarray}}
\def\eeqa{\end{eqnarray}}
\def\bean{\begin{eqnarray*}}
\def\eean{\end{eqnarray*}}

\renewenvironment{thebibliography}[1]
         {\section*{References}\frenchspacing\small
          \begin{list}{[\arabic{enumi}]}
         {\usecounter{enumi}\parsep=2pt\topsep 0pt
         \settowidth{\labelwidth}{[#1]}
         \leftmargin=\labelwidth\advance\leftmargin\labelsep
         \rightmargin=0pt\itemsep=1pt\sloppy}}{\end{list}}

 \numberwithin{equation}{section}

\input xy
\xyoption{all}

\usepackage[normalem]{ulem}

\newcommand{\ket}[1]{\left| #1 \right\rangle}

\newcommand{\vev}[1]{\left\langle #1 \right\rangle}

\newcommand{\braket}[2]{\left\langle \vphantom {#1 #2} #1 \hphantom{|} \right| \left. \vphantom {#1 #2} #2 \right\rangle}

\newcommand{\braopket}[3]{\left\langle \vphantom {#1 #2 #3} #1 \hphantom{|} \right| #2 \left| \hphantom{|} \vphantom {#1 #2 #3} #3 \right\rangle}

\newcommand{\ba}{\begin{eqnarray}}
\newcommand{\ea}{\end{eqnarray}}

\title{\textbf{\textsf{Coarse graining as a representation change}}\vspace{0.35cm}}

\author{
\textsf{Norbert Bodendorfer\affmark[1]\footnote{\texttt{norbert.bodendorfer@physik.uni-r.de}}, Fabian Haneder\affmark[1]\footnote{\texttt{fabian.haneder@stud.uni-regensburg.de}}}\\
\affaddr{\affmark[1]\textsf{Institute for Theoretical Physics, University of Regensburg,}}\\
\affaddr{\textsf{93040 Regensburg, Germany}}\vspace{-0.5cm}
}

\begin{document}

\maketitle

\begin{abstract}
\textsf{We discuss how SU$(1,1)$ coherent states from the discrete series allow for a natural coarse graining operation. The physical application are quantum theories based on a set of three extensive observables whose Poisson algebra is isomorphic to su$(1,1)$. In particular, we show that a Perelomov coherent state with representation label $N j_0$ and spinor label $z$ encodes the physics of $N$ independent subsystems with labels $j_0, z$. This property is suggested by existing results for the expectation values and variances of the observables. We prove that it holds for all higher moments. Our results in particular apply to a recent quantum cosmology model that has been derived using SU(1,1) group theoretic quantisation techniques. For it, it follows that a certain notion of fiducial cell independence holds exactly at the quantum level when using the coherent states. }
\end{abstract}

\section{Introduction}

Coarse graining is a notoriously difficult problem, yet of fundamental importance if one wants to understand large scale features of a system with a large number of microscopic constituents. While a purely analytic understanding is usually out of the question, one is still interested in toy models which can be coarse grained exactly. Such toy models may be applicable to certain physical situations that are sufficiently symmetric, such as the following one. 

The physical setting that lead to this paper is quantum cosmology \cite{CalcagniBook}. Already in classical cosmology, one works with fiducial cells that capture only a finite part of the (possibly infinitely large) universe, but gladly accepts this tradeoff due to the homogeneity assumption. It is then of importance to establish that the final results do not depend on this choice of fiducial cell, which can be interpreted as a scale change in the system. While fiducial cell independence usually holds in classical cosmology, the situation is different for quantum cosmology due to the existence of a new quantum gravity scale. While fiducial cell independence is expected to hold for large cells, see e.g. \cite{CorichiOnTheSemiclassical} for an explicit confirmation, it is unclear what happens as the proper fiducial cell size approaches the quantum gravity scale. A general recent discussion can e.g. be found in \cite{BojowaldTheBKLScenario}. 

Given this situation, it would be interesting to know whether there exists a theory of quantum cosmology with a set of quantum states that can be coarse grained exactly and whether fiducial cell independence holds. In this paper, we will show that a recently proposed class of models \cite{LivineGroupTheoreticalQuantization, BenAchourThiemannComplexifierIn} using SU$(1,1)$ group theoretic quantisation techniques qualifies. In particular, the representation label $j\in \mathbb N/2$ of SU$(1,1)$ Perelomov coherent states \cite{PerelomovCoherentStatesFor} serves as a scale of the system. We will show that the physics (encoded in the expectation values of the three observables and their higher moments) of $N$ independent cells with coherent state label $\ket{j_0, z}$, $z \in \mathbb C^2$, is fully captured by a coherent state with label $\ket{N j_0, z}$. This in particular requires to compare the scaling of moments with $N$ to an analogous computation using beyond-Gau{\ss}ian error propagation. Our computation builds on the earlier suggestion \cite{BodendorferStateRefinementsAnd}, which was based on a quantum cosmology model \cite{AshtekarRobustnessOfKey} where fiducial cell independence could only be argued for large cells using the results of \cite{CorichiOnTheSemiclassical}. We note that the recent work \cite{BenAchourScaleInvarianceWith} discusses a related, but conceptually different notion of scale invariance in the cosmological model \cite{BenAchourThiemannComplexifierIn}, which also rescales an intensive quantity (the Hubble rate $b$). We will make our concept of fiducial cell independence more precise in the conclusion section, as it requires some notions introduced in the main text.

Our result turns out to be quite general despite the intended application to quantum cosmology. We only require to consider three real observables, including the generator of time translations, that scale extensively and whose Poisson algebra is isomorphic to su$(1,1)$.

\section{Group quantisation with SU(1,1)}

In the following, we will recall the idea of quantising a given physical system by identifying the classical Poisson algebra with the Lie algebra of a group whose representation theory is known, specifically SU$(1,1)$, see \cite{IshamTopologicalAndGlobal} for a seminal contribution. We will remain rather abstract in this paper as our conclusions do not rely on the specific physical interpretation of the algebra elements. 
As an example, one may consider the quantum cosmological models discussed in \cite{LivineGroupTheoreticalQuantization, BenAchourThiemannComplexifierIn}, see also \cite{BojowaldDynamicalCoherentStates, BojraDynamicsForA} for earlier research connecting SU$(1,1)$ with quantum cosmology. 

We start with three phase space functions $j_z$, $j_+$, and $j_-$ that satisfy the algebra
\be
	\{j_z, j_\pm\} = \mp i j_\pm, ~~~~ \{j_+, j_-\} =  2 i j_z \text{.}
\ee
They are considered to be extensive observables, i.e. scale linearly in the volume of the system. Such quantities may be obtained by integrating densities over space, e.g. the volume and Hamiltonian densities in a cosmological model as in \cite{LivineGroupTheoreticalQuantization, BenAchourThiemannComplexifierIn}. Reality conditions are such that $j_+ = \overline {j_-}$ and $j_z = \overline {j_z}$. Alternatively, we could have considered the phase space functions $j_z$, $j_x$, and $j_y$ with $j_\pm=j_x\pm i j_y$, so that $j_z$, $j_x$, and $j_y$ are all real and extensive. 

Since our sole interest will be in those functions, we obtain the respective quantum theory by finding a representation of the Poisson algebra on a Hilbert space via the quantisation rule $[\cdot, \cdot] = i \{ \cdot, \cdot\}$ with $\hbar = 1$. We are thus looking for linear operators  $\hat j_z$, $\hat j_+$, and $\hat j_-$ satisfying 
\be
	[\hat j_z, \hat j_\pm] = \pm \hat j_\pm, ~~~~ [\hat j_+, \hat j_-] =  - 2  \hat j_z \text{,}
\ee
which is equivalent to studying the representation theory of the Lie algebra su$(1,1)$. 

As an example, the defining representation of su$(1,1)$ yields the operators
\be
	\hat j_z = \sigma_z , ~~~~~ \hat j_+= \sigma_x + i \sigma_y, ~~~~~ \hat j_-= \sigma_x - i \sigma_y
\ee
with
\be
	\sigma_z = \frac{1}{2}  \begin{pmatrix} 
1 & 0 \\
0 & -1 
\end{pmatrix}, ~~~ \sigma_x = \frac{1}{2}  \begin{pmatrix} 
0 & 1 \\
-1 & 0 
\end{pmatrix}, ~~~\sigma_y = \frac{-i}{2}  \begin{pmatrix} 
0 & 1 \\
1 & 0 
\end{pmatrix} \text{.}
\ee
They act on a two-dimensional complex space which we coordinatise with the two complex numbers $z^0$, $z^1 \in \mathbb C$ that we combine into single spinor $z = \begin{pmatrix} 
z^0\\ z^1 \end{pmatrix}$. By construction, a finite transformation $U = \exp(i \alpha^k \sigma_k)$, $k = z, x, y$, acting on spinors as $U \cdot z$ preserves $(\tilde z)^\dagger \cdot \epsilon \cdot z$ with $\epsilon= \begin{pmatrix} 
1 & 0 \\
0 & -1 
\end{pmatrix}$. Note that since SU$(1,1)$ is non-compact, this cannot be a unitary representation, e.g. w.r.t. the standard (positive definite) scalar product $(\tilde z)^\dagger \cdot z$. 

Rather, unitary irreducible representations of SU$(1,1)$ are necessarily infinite-dimensional, see e.g. \cite{RamondBook} for an overview. In this paper, we will be interested in the representations from the discrete class with representation label (spin) $j \in \mathbb N/2$. States in a representation space with spin $j$ are labelled by magnetic indices $m = j, j+1, j+2, \ldots$\footnote{There are also analogous representations negative $m$ that we will not consider here, as well as representations with $j=\frac14$ or $j=\frac34$, and two continuous classes.}. Their scalar product reads $\braket{j, m}{j', m'}=\delta_{j, j'} \delta_{m, m'}$.

We will not concern us with the precise action of $\hat j_z$, $\hat j_+$, and $\hat j_-$ on these representation spaces as a suitable choice of coherent states will allow us in the next section to reduce the effective dimension of a representation space with label $j$ to that of the above representation on spinors. This concludes our brief exposition of what one may call ``group theoretic quantisation'' as e.g. in \cite{LivineGroupTheoreticalQuantization}.

\section{Coarse graining}

\subsection{SU(1,1) coherent states, expectation values, and variances} 

Abbreviating $L = \frac12 (|z^0|^2-|z^1|^2)$, we define the usual {\it normalised} SU$(1,1)$ coherent states \cite{PerelomovCoherentStatesFor}, 
\be
	\ket{j, z} = (2L)^{j} \sum_{m=j}^\infty \sqrt{\binom{m+j-1}{m-j}} \, \frac{(z^1)^{m-j}}{(\bar{z}^0)^{m+j}} \, \ket{j,m} \label{eq:DefCoh}
\ee
In the following, we will go through their properties and link them to the desired coarse graining interpretation. We restrict the spinor labels so that $L>0$, which is preserved by the SU$(1,1)$ action. 

Let us first note the property that an SU$(1,1)$ transformation acts only on the spinors (e.g. \cite{LivineGroupTheoreticalQuantization}), i.e. 
\be
	U\ket{j,z} = \ket{j, U \cdot z} ~ \forall ~ U \in \text{SU}(1,1) \text{,}
\ee
which allows an important conclusion: when working with these coherent states, the effective dimension of the Hilbert space is finite dimensional and labelled by the two complex numbers $z^0$ and $z^1$. In particular, the dynamics, which is generated by an element of su$(1,1)$ (we assume the generator of time translations to be a linear combination of $j_z$, $j_+$, and $j_-$), leaves this subspace invariant. It should also be noted that the representation label $j$ is invariant under the dynamics. 

Next, we recall the expectation values of the three basic operators $\hat j_z$, $\hat j_+$, and $\hat j_-$ (e.g. \cite{LivineGroupTheoreticalQuantization}):
\ba
	\braopket{j, z}{\hat j_z}{j, z} &= &  j \frac{|z^0|^2+|z^1|^2}{2L} := j \frac{v_z}{L} \\
	\braopket{j, z}{\hat j_+}{j, z} &= & j \frac{\bar{z}^0 \bar{z}^1}{L}  :=  j \frac{v_+}{L} \\
	\braopket{j, z}{\hat j_-}{j, z} &= &  j \frac{z^0 z^1}{L} := j \frac{v_-}{L} 
\ea
Their scaling with $j$ suggests to interpret $j$ as a scale of the system that decouples from the scale-invariant state determined by $v_z$, $v_+$, $v_-$, and $L$, i.e. the spinors.

For this interpretation to be tenable, we need to also check the scaling of the variances as a function of $j$ (e.g. \cite{LivineGroupTheoreticalQuantization}): 
\ba
	\sigma_z^2 = \braopket{j, z}{\hat j_z^2}{j, z}-\braopket{j, z}{\hat j_z}{j, z}^2 &= &  \frac{j}{2} \left( \frac{v_z^2}{L^2}-1 \right) \label{eq:VarZ}\\
	\sigma_+^2= \braopket{j, z}{\hat j_+^2}{j, z}-\braopket{j, z}{\hat j_+}{j, z}^2 &= & \frac{j}{2} \left( \frac{v_+^2}{L^2} +1\right) \label{eq:VarP} \\
	\sigma_-^2= \braopket{j, z}{\hat j_-^2}{j, z}-\braopket{j, z}{\hat j_-}{j, z}^2&= &  \frac{j}{2}  \left( \frac{v_-^2}{L^2} +1 \right) \label{eq:VarM}
\ea
Standard Gau{\ss}ian error propagation of uncorrelated systems tells us that variances add up, i.e. the variance of a coarse grained system with representation label $j$ is equal to $j$ times the variance at $j=1$, which is consistent with the linear scaling in \eqref{eq:VarZ}-\eqref{eq:VarM}. The scaling in e.g. $v_z$ on the other hand shows that interpreting a change in the spinors as a coarse grained scale change does not comply with error propagation. We also note that the relative errors scale as $\frac{\sigma_\alpha}{\vev{j_\alpha} }\propto \frac{1}{\sqrt{j}}$, $\alpha = z, +, -$, as a function of $j$, showing that the system behaves more and more classical as $j$ increases. On the other hand, relative errors do not scale with the spinors for large $v_\alpha/L$, showing that the system  does not behave more and more classical as we zoom out by increasing the extensive quantities via the spinors. 

These considerations support the above interpretation of $j$ as a scale, while the spinor $z$ sets the ``intensive'' quantum state of individual quanta in the system. This observation suggests the coarse graining interpretation advocated here, but it still may be that a coherent state with $j>1$ does not fully capture the physics of $j$ uncorrelated systems at $j=1$ in that the higher moments $\Delta_{\alpha}^n := \vev{ \left(\hat j_\alpha-\vev{\hat j_\alpha}\right)^n}$, $n\geq 3$, could be different. Surprisingly, this is not the case as we will show in the next section. 

For pedagogical reasons, we chose to first explain our findings when coarse graining fundamental systems with $j=1$. We will see however that we may choose any fundamental $j_0 \in \mathbb N/2$.

\subsection{Beyond-Gau{\ss}ian propagation of errors}

Let us first establish the short hand notation $\vev{\hat j_\alpha^n}_j= \braopket{j,z}{\hat j_\alpha^n}{j,z}$.
For the higher moments $\Delta_{\alpha}^n := \vev{ \left(\hat j_\alpha-\vev{\hat j_\alpha}_j\right)^n}_j$, we abbreviate $q=\frac{|z^1|^2}{|z^0|^2}<1$ and first compute
\ba
	\vev{\hat j^n_z}_j &=&(1-q)^{2j} \sum_{k=0}^\infty \binom{2j+k-1}{k} q^k (k+j)^n \label{eq:jzPower}\\
	  &=&(1-q)^{2j} \sum_{i=0}^n \binom{n}{i} j^{n-i} \left( q \frac{\partial}{\partial q}\right)^i \frac{1}{(1-q)^{2j}}
\ea
and
\ba
	\vev{\hat j^n_+}_j  &=& (1-q)^{2j}\left( \frac{\bar{z}^1}{z^0}\right)^n \sum_{k=0}^{\infty} \binom{2j+k-1}{k} q^k \frac{(2j+k-1+n)!}{(2j+k-1)!} \\
	 &=&(1-q)^{2j}\left( \frac{\bar{z}^1}{z^0}\right)^n \sum_{i=0}^{n}  \left( q \frac{\partial}{\partial q}\right)^i \frac{1}{(1-q)^{2j}}  \frac{1}{i!} \left. \left( \frac{\partial}{\partial k} \right)^i \right|_{k=0} \frac{(2j+k-1+n)!}{(2j+k-1)!} \text{.} ~~~
\ea
Furthermore, $\vev{\hat j^n_-}_j = \overline{\vev{\hat j^n_+}}_j$. 
The moments then follow as 
\be
	\Delta_{\alpha}^n = \sum_{i=0}^n \binom{n}{i} (-1)^i \vev{\hat j_\alpha^{n-i} }_j\vev{\hat j_\alpha }_j^i \text{.}
\ee

To compare the $j$-dependence of the higher moments with beyond-Gau{\ss}ian error propagation, we need to discuss how higher moments behave in joint uncorrelated measurements. We consider $j$ independent identical ensembles with a random variable $x$ and denote them by $x_i$, $i = 1, \ldots, j$ with expectation value functional $\vev{\cdot}$ (a priori different from $\braopket{j,z}{\cdot}{j,z}$). 
We are interested in the coarse grained observable $X := \sum_{i=1}^j x_i$. Standard Gau{\ss}ian error propagation follows from the independence assumption as 
\be
	\sigma_X^2 =  \vev{\left( \sum_{i=1}^j \left(x_i - \vev{x_i} \right) \right)^2  }=\sum_{i=1}^j  \vev{ \left(   x_i  -\vev{x_i}  \right) \left(    x_i  -\vev{x_i} \right) } = j \vev{\left(x-\vev{x} \right)^2}= j \sigma_x^2 \text{,}
\ee
i.e. products of two brackets where the index $i$ is different vanish due to $\vev{ x_i - \vev{x_i} }=0$. For higher powers $n > 2$, the derivation proceeds similarly: from the $n$ sums with $n$ independent summation indices $i_a$, $a=1, \ldots, n$, terms will drop by the independence assumption unless every index $i_a$ appears at least twice in the product. One then needs to sum over all possibilities how this may happen and pick the correct combinatorial prefactors. The result can be obtained using the multinomial theorem and conveniently rewritten as 
\ba
	 \vev{\left(X-\vev{X} \right)^n} \label{eq:NGEP}
	 &=&   \sum_{\substack{ r_1, \ldots, r_{j} = 0  :\\ n = r_1 + \ldots + r_{j}}}^n \frac{n!}{r_1! r_2! \ldots r_{j}!} \vev{ (x_1 - \vev{x_1})^{r_1}} \ldots \vev{( x_j - \vev{x_j} )^{r_j}}  \\ 
	 & = & \sum_{m=1}^{\floor*{n/2}} \frac{j!}{(j-m)!} \sum_{\substack{2 \leq k_1 \leq \ldots \leq k_{m}  :\\ n = k_1 + \ldots + k_{m}}} \frac{n!}{k_1! k_2! \ldots k_{m}!} \prod_{p=2}^{n} \frac{1}{(\#k_i=p)!} \vev{\left(x-\vev{x}\right)^p}^{(\#k_i=p)}  \text{,} \nonumber
\ea 
where in the last line, $(\#k_i=p)$ is the number of times an index $k_i$ takes the value $p$. The index $m$ summed over first counts how many different values the indices $i_a$ take. The first coefficient counts the number of possibilities we can pick $m$ different indices from all summation indices $i_a$, accounting also for the picking order. All these choices lead to the same term due to the subsystems being identical. 
The second and third terms count the number of possibilities that a certain partitioning $n = k_1 + \ldots + k_{m}$ with $2 \leq k_i \leq k_{i+1}$ can occur, i.e. how many of the $n$ brackets have the same indices $i_1, \ldots, i_m$, $i_i \neq i_{\tilde i}$, $i, \tilde i=1, \ldots, m$, with the third term accounting for a double-counting when some of the $k_i$ agree. 
The upper bound $\floor*{n/2}$ for $m$ is obtained for even $n$ as $n = 2 + 2 + \ldots +2 = 2 m$ and odd $n$ as $n = 2 + 2 + \ldots + 2+3 = 2(m-1)+3$.

As an example, consider $n=4$. $m=1$ gives a term linear in $j$, whereas the partitioning $4=2+2$ for $m=2$ yields a non-trivial correction to this linear scaling as
\ba
	\vev{\left(X-\vev{X} \right)^4} = j \vev{\left(x-\vev{x} \right)^4} +3 j (j-1) \vev{\left(x-\vev{x} \right)^2}^2 \text{.}
\ea
The combinatorial factor $3$ accounts for the choices $()_{i_1}()_{i_1}()_{i_2}()_{i_2}$, $()_{i_1}()_{i_2}()_{i_1}()_{i_2}$, and $()_{i_1}()_{i_2}()_{i_2}()_{i_1}$ once the indices $i_1$ and $i_2$ are picked, taking into account the order of the picking, which gives another factor of $j(j-1)$. 

We are now in a position to compare this result for beyond-Gau{\ss}ian error propagation to the the $j$-dependence of the coherent states. As said before, we consider $j$ as a label for the coarse graining of $j$ independent identical subsystems. This leads to the claim
\ba
	&& \vev{\left(\hat j_\alpha-\vev{\hat j_\alpha}_j \right)^n}_j \label{eq:Conjecture}\\
	&=&  \sum_{m=1}^{\floor*{n/2}} \frac{j!}{(j-m)!} \sum_{\substack{2 \leq k_1 \leq \ldots \leq k_{m}  :\\ n = k_1 + \ldots + k_{m}}} \frac{n!}{k_1! k_2! \ldots k_{m}!} \prod_{p=2}^{n} \frac{1}{(\#k_i=p)!} \vev{\left(\hat j_\alpha-\vev{\hat j_\alpha }_1 \right)^p}_1^{(\#k_i=p)} \nonumber
\ea
which means that the coherent states with label $j,z$ encode all moments of the probability distribution coming from considering $j$ independent systems at level $j=1$ with the same spinor label $z$. Since the dynamics of the system is generated by one of the three generators that acts only on the spinors, the dynamics of the coarse grained systems also agrees with the fundamental dynamics at $j=1$. 

Before continuing, consider again $n=4$ as an example and $\alpha = z$. We obtain
\ba
	&& \vev{\left(\hat j_z-\vev{\hat j_z}_j \right)^4}_j \\
	&=& \nonumber
	2j \frac{|z^0|^2|z^1|^2}{(|z^0|^2-|z^1|^2)^2} \left( 1+6(j+1) \frac{|z^0|^2|z^1|^2}{(|z^0|^2-|z^1|^2)^2}  \right) \\
	&=&j  \underbrace{2 \frac{|z^0|^2|z^1|^2}{(|z^0|^2-|z^1|^2)^2}   \frac{|z^0|^4+ |z^1|^4+10 |z^0|^2|z^1|^2}{(|z^0|^2-|z^1|^2)^2}}_{ \vev{\left(\hat j_z-\vev{\hat j_z}_1 \right)^4}_1}  + 3j(j-1) \underbrace{\left(2 \frac{|z^0|^2|z^1|^2}{(|z^0|^2-|z^1|^2)^2}\right)^2}_{\vev{\left(\hat j_z-\vev{\hat j_z}_1 \right)^2}_1^2} \nonumber
\ea
from both computations, and thus agreement.

\subsection{General form of results}

So far, we have used $j=1$ as a representation label for the subsystems to be coarse grained for pedagogical reasons. However, any $j\in \mathbb N/2$ can be used.
Let us therefore start with a set of $N \in \mathbb N$ identical independent systems with representation label $j_0$. We claim that their dynamics is fully captured in the above sense by a coherent state with representation label $j = N j_0$ and the same spinor labels as
\ba
	&& \vev{\left(\hat j_\alpha-\vev{\hat j_\alpha}_j \right)^n}_j \label{eq:ConjectureFull}\\
	&=&  \sum_{m=1}^{\floor*{n/2}} \frac{N!}{(N-m)!} \sum_{\substack{2 \leq k_1 \leq \ldots \leq k_{m}  :\\ n = k_1 + \ldots + k_{m}}} \frac{n!}{k_1! k_2! \ldots k_{m}!} \prod_{p=2}^{n} \frac{1}{(\#k_i=p)!} \vev{\left(\hat j_\alpha-\vev{\hat j_\alpha }_{j_0} \right)^p}_{j_0}^{(\#k_i=p)} \text{.} \nonumber
\ea
Alternatively, similar arguments as above lead to
\ba
	 \vev{\hat j_\alpha ^n}_j \label{eq:ConjectureFullPP} 
	&=& \sum_{\substack{ r_1, \ldots, r_{j} = 0  :\\ n = r_1 + \ldots + r_{j}}}^n \frac{n!}{r_1! r_2! \ldots r_{j}!} \vev{ \hat j_\alpha^{r_1}}_{j_0} \ldots \vev{\hat j_\alpha^{r_j}}_{j_0}  \\    
	&=& \sum_{m=1}^{n} \frac{N!}{(N-m)!} \sum_{\substack{1 \leq k_1 \leq \ldots \leq k_{m}  :\\ n = k_1 + \ldots + k_{m}}} \frac{n!}{k_1! k_2! \ldots k_{m}!} \prod_{p=1}^{n} \frac{1}{(\#k_i=p)!} \vev{\hat j_\alpha^p}_{j_0}^{(\#k_i=p)} \text{,} 
\ea
where we note that $k_1, \ldots, k_m$ now may take the value $1$ in addition, the first sum runs until $m=n$, and the product starts at $p=1$. Since one can extract the coarse grained moments \eqref{eq:ConjectureFull} and expectation values of powers \eqref{eq:ConjectureFullPP} from each other, both statements are equivalent. We prove \eqref{eq:ConjectureFullPP} in the appendix.

\subsection{Eigenvalues and their probabilities}

In addition to the expectation values, we may also be interested how the eigenvalues and the probabilities to obtain them behave under coarse graining. We consider $\alpha = z$. For the representation $j_0$, the eigenvalues of $\hat j_z$ are $j_0 + k$, $k = 0, 1, 2, \ldots$. After combining $N$ such systems, the possible eigenvalues of the coarse observable are $N j_0 + k$, $k = 0, 1, 2, \ldots$, which agrees with the possible eigenvalues of $\hat j_z$ in the representation $N j_0$. 

Even more can be said: from \eqref{eq:DefCoh}, we obtain the probability to measure the eigenvalue $j_0+k$ with eigenvector $\ket{j_0,k}$ in the representation $j_0$ as
\be
	P_{j_0,k} := \left| \braket{j_0, k}{j_0, z}\right|^2 = (1-q)^{2j_0} q^k \binom{2j_0+k-1}{k} \text{.}
\ee
The probability to measure the coarse grained eigenvalue $Nj_0+k$ is obtained as
\be
	P^{\text{coarse}}_{Nj_0,k} = \sum_{\substack{ k_1, \ldots, k_{N} = 0  :\\ k = k_1 + \ldots + k_{N}}} P_{j_0,k_1} \cdot P_{j_0,k_2} \cdot  \ldots  \cdot P_{j_0,k_N} \text{.} \label{eq:ProbPropagation}
\ee
We show in the appendix that $P^{\text{coarse}}_{Nj_0,k} = P_{Nj_0,k}$, i.e. the coarse graining operation also captures the correct probabilities for the eigenvalues of $\hat j_z$.

\section{Conclusion}

We have demonstrated that SU$(1,1)$ coherent states have a natural coarse graining interpretation when rescaling the representation label $j$. This allows for an interesting application to quantum cosmology based on \cite{LivineGroupTheoreticalQuantization, BenAchourThiemannComplexifierIn}, where one can capture the coarse grained dynamics of $N$ non-interacting subsystems described by Perelomov coherent states with representation label $j_0$ by a single Perelomov coherent state with label $Nj_0$. Our results hold for the time evolution of the expectation values of the observables and their higher moments, but we have not shown that all possible quantum properties of the states are captured.

For loop quantum cosmology, the present computation applied to \cite{LivineGroupTheoreticalQuantization, BenAchourThiemannComplexifierIn} gives the first example where fiducial cell independence can be established exactly. 
The precise notion we use here is as follows. We fix the total {\it physical} spatial volume as a comparison of different physical spatial volumes means comparing different physical systems. The total volume is split into $N$ identical cells which function as fiducial cells. Within each cell, we consider the three observables corresponding to the su$(1,1)$ generators. The coarse grained observables are defined as the sums of the cell observables. Initial conditions are set (identically in each cell) by specifying the spinor $z$ and the representation $j$ defining \eqref{eq:DefCoh}. Evolution is generated by a linear combination of the observables. Our results show that instead of considering $N$ such identical fiducial cells, we could have equally well considered a single cell with representation label $Nj$ and the same spinor label to compute the expectation values of all moments of the three observables, including the correct dynamics. We note that it was crucial to correctly use Beyond-Gau{\ss}ian error propagation to get the correct behaviour of the moments.
It should be specifically emphasised that while the relative variances increase when lowering the cell size as $j \rightarrow j/N$, this effect is exactly cancelled by considering a collection of $N$ such cells and computing the coarse grained variances. 

In the context of loop quantum gravity where \cite{LivineGroupTheoreticalQuantization, BenAchourThiemannComplexifierIn} is situated, this provides the first example of a simplified model where it can be established that the collective dynamics of many (arbitrarily) low spins (representations) agrees (exactly) with the large spins dynamics. In a broader context, this is an important step in establishing the classical limit of the theory for quantum states with many small instead of a few large quantum numbers. While the discussion in this paper was limited to quantum cosmology, one can also view the computation here as being embedded into full quantum gravity as discussed in \cite{BVI}. Hence, it is of obvious interest for future research to determine to which extent it can be generalised beyond the homogeneous and isotropic setting.

\section*{Acknowledgments}

NB was supported by an International Junior Research Group grant of the Elite Network of Bavaria.

\newpage

\begin{appendix}

\section{Proofs}

For the proofs, it is key to repeatedly use the identity 
\be
	\sum_{i=0}^k \binom{a+i-1}{i}\binom{b+k-i-1}{k-i} = \binom{a+b+k-1}{k} \text{.} \label{eq:CoreIdentity}
\ee

We outline the proof for the expectation value $\vev{\hat j_z^n}$ in detail. The proof for $\vev{\hat j_+^n}$ is analogous and requires an additional step sketched below. Coarse graining $N$ identical systems leads to 
\ba
	\vev{\hat j_z^n}^{\text{coarse}} &=& \sum_{\substack{ r_1, \ldots, r_{N} = 0  :\\ n = r_1 + \ldots + r_{N}}} \frac{n!}{r_1! \ldots r_N!}  \vev{ \hat j_\alpha^{r_1}}_{j_0} \ldots \vev{\hat j_\alpha^{r_j}}_{j_0} \label{eq:Proof} \\
	&=&(1-q)^{2Nj_0} \sum_{\substack{ r_1, \ldots, r_{N} = 0  :\\ n = r_1 + \ldots + r_{N}}} \frac{n!}{r_1! \ldots r_N!}  \left( \sum_{i_1=0}^\infty \binom{2j_0+i_1-1}{i_1} q^{i_1} (i_1+2j_0)^{r_1}\right) \nonumber  \\
	&&~~~~~~~~~~~~~~~~~~~~~~~~~~~~~~~~~~~~~~~~ \times \ldots \times  \left( \sum_{i_N=0}^\infty \binom{2j_0+i_N-1}{i_N} q^{i_N} (i_N+2j_0)^{r_N}\right) \nonumber \\
	&=&(1-q)^{2Nj_0} \sum_{i_1, \ldots, i_N=0}^\infty q^{\sum_{k=1}^N i_k}  \binom{2j_0+i_1-1}{i_1}  \ldots \binom{2j_0+i_N-1}{i_N} \left({\sum_{k=1}^N i_k}+2 N j_0 \right)^n \nonumber\\
	&=&(1-q)^{2Nj_0} \sum_{k_1=0}^\infty \sum_{k_2=0}^{k_1} \ldots \sum_{k_N=0}^{k_{N-1}} q^{k_1}  \binom{2j_0+k_N-1}{k_N}  \binom{2j_0+k_{N-1}-k_N-1}{k_{N-1}-k_N} \nonumber \\
	&&~~~~~~~~~~~~~~~~~~~~~~~~~~~~~~~~~~~~~~~~~~~~~~~ \times  \ldots \times \binom{2j_0+k_{1}-k_2-1}{k_{1}-k_2}      \left(k_1+2 N j_0 \right)^n \nonumber\\
	&=&(1-q)^{2Nj_0} \sum_{k_1=0}^\infty q^{k_1}  \binom{2 N j_0+k_1-1}{k_1}  q^{k_1} (k_1+N j_0)^n \nonumber \\
	&=& \vev{\hat j_z^n}_{N j_0} \nonumber
\ea
In the first line, we used the coarse graining \eqref{eq:ConjectureFullPP} for $N$ independent subsystems. In the second line, we inserted the expectation values for representation label $j_0$. In the third line, we used the multinomial theorem to perform the sum over $r_1, \ldots, r_N$. In the fourth line, we rewrote the expression using the Cauchy product for multiple series. The fifth line is obtained by applying \eqref{eq:CoreIdentity} $N-1$ times. Renaming $k_1 \rightarrow k$ yields \eqref{eq:jzPower} for $j = Nj_0$ and the claim follows. 

The proof for $\vev{\hat j_\pm^n}$ proceeds analogously. Instead of the multinomial theorem, we have to use 
\ba
	&& \sum_{\substack{ r_1, \ldots, r_{N} = 0  :\\ n = r_1 + \ldots + r_{N}}} \frac{n!}{r_1! \ldots r_N!} \frac{(2j_0+i_1-1+r_1)!}{(2j_0+i_1-1)!} \ldots \frac{(2j_0+i_N-1+r_N)!}{(2j_0+i_N-1)!}\\
	&=& \frac{(2 N j_0+ \sum_{k=1}^N i_k  -1+n)!}{(2Nj_0+\sum_{k=1}^N i_k -1)!} \text{,} \nonumber 
\ea
which again follows from a repeated application of \eqref{eq:CoreIdentity}.
The proof of \eqref{eq:ProbPropagation} is a simplified version of \eqref{eq:Proof} and consists of the steps from lines three to five.

\end{appendix}

%\bibliographystyle{utphysmendeley}
%\bibliography{library}

\end{document}